\documentclass[11pt]{article}

\usepackage[utf8]{inputenc}
\usepackage[T1]{fontenc}
\usepackage{lmodern}
\usepackage{amsmath, amssymb, amsfonts}
\usepackage{booktabs}
\usepackage{float}
\usepackage[margin=1in]{geometry}
\usepackage{hyperref}
\usepackage{natbib}
\usepackage{graphicx}
\usepackage{txfonts}
\usepackage{lineno}

\title{Systemic Risk and Default Cascades in Global Equity Markets: A Network and Tail-Risk Approach Based on the Gai Kapadia Framework}

\author{
Ana I. Castillo Pereda\\
\small Institute of Mathematics and Statistics, University of São Paulo, São Paulo, SP, Brazil\\
\small \textit{Correspondence:} anacp20@gmail.com; anaicp@ime.usp.br\\
\small Tel.: +55-11-983967981
}

\date{April 2026}

\begin{document}

\maketitle
\begin{abstract}
This study extends the Gai–Kapadia framework, originally developed for interbank contagion, to assess systemic risk and default cascades in global equity markets. We analyze a 30-asset network comprising Brazilian and developed-market equities over the period 2015–2026, constructing exposure-based financial networks from price co-movements. Threshold filtering ($\theta = 0.3$ and $\theta = 0.5$) is applied to isolate significant interconnections.

Cascade dynamics are analyzed through a combination of deterministic propagation and stochastic Monte Carlo simulations ($n = 1000$) under varying shock intensities. The results show that the system exhibits strong global resilience, with a negligible probability of large-scale failure, while maintaining localized vulnerability within highly clustered subnetworks. In particular, shocks lead to an average of 1.0 failed asset for single shocks and 2.0 for simultaneous shocks, indicating limited propagation below a critical threshold.

Network analysis reveals a clear structural asymmetry: Brazilian assets display high clustering ($C_i \approx 0.8$-$1.0$) and dense connectivity, which amplifies local shock propagation, whereas developed-market assets exhibit lower connectivity ($C_i \approx 0.2$-$0.5$), limiting systemic spread. Tail risk analysis, based on empirical CCDF and Hill estimators, confirms the presence of heavy-tailed loss distributions, particularly in emerging markets, reinforcing their exposure to extreme events.

These findings demonstrate that systemic risk arises from the interaction between network topology and tail behavior, rather than from isolated asset characteristics. The proposed framework provides a scalable and empirically grounded approach for stress testing and systemic risk assessment, offering relevant insights for regulators and portfolio managers in increasingly interconnected financial markets.

\noindent\textbf{Keywords:}
Systemic risk; Default cascades; Financial networks; Equity markets; 
Network contagion; Tail risk; Heavy-tailed distributions; 
Monte Carlo simulations; Value-at-Risk (VaR); Conditional Value-at-Risk (CVaR)

\end{abstract}
\vspace{0.1cm}

\section{Introduction}

Systemic risk and default cascades pose significant threats to financial stability, as interconnected assets can amplify shocks across global equity markets. This study extends the Gai Kapadia framework \cite{Gai2010}, originally developed for interbank contagion, to quantify systemic risk in equity markets using price co-movements, exposure networks, and stochastic simulations. We analyze a network of 30 assets (15 Brazilian and 15 developed-market equities) over the period 2015 - 2026, constructing exposure-based networks under thresholds \(\theta \in \{0.3, 0.5\}\) and evaluating cascade dynamics through Monte Carlo simulations (\(n = 1000\)) with shocks ranging from 10\% to 50\%.

Traditional risk measures such as Value-at-Risk (VaR) and Conditional Value-at-Risk (CVaR) provide important insights into downside risk but often fail to capture network-driven contagion effects \cite{MantegnaStanley1999, Eisenberg2001}. This limitation motivates the use of graph-theoretic approaches, which explicitly account for interdependencies among assets \cite{Newman2010, Strogatz2001, Poledna2021, Ellis2022}. In this context, financial networks offer a natural framework for modeling how shocks propagate through interconnected systems.

While the Gai Kapadia framework provides a robust foundation for modeling systemic risk through exposure networks, its application to equity markets remains relatively underexplored \cite{Glasserman2016}. Our study contributes to this literature by integrating correlation-based networks, exposure dynamics, and stochastic simulations, providing a unified approach to modeling contagion in equity markets. Building on prior work by Acemoglu et al. \cite{Acemoglu2015} and Battiston et al. \cite{Battiston2012}, we analyze how network structure influences systemic stability, extending systemic risk measures proposed by Billio et al. \cite{Billio2012} and Haldane and May \cite{Haldane2011}.

A key contribution of this study is the integration of tail risk into the network framework. Empirical evidence shows that financial returns exhibit heavy-tailed behavior, implying a higher likelihood of extreme events than predicted by Gaussian models \cite{Cont2001, Mandelbrot1963}. By combining network topology with tail risk measures, this study provides a more comprehensive understanding of systemic vulnerability, where highly connected assets with heavy-tailed loss distributions act as potential amplifiers of contagion.

Recent geopolitical and macroeconomic disruptions further underscore the importance of systemic risk modeling in globally interconnected markets. Shocks originating in specific sectors such as energy supply disruptions or trade fragmentation can propagate across financial systems, reinforcing the need for models that capture cross-market contagion mechanisms. These real-world dynamics align with the network-based perspective adopted in this study, where localized shocks can generate broader systemic effects.

The contributions of this study are as follows:
\begin{itemize}
    \item Extension of the Gai Kapadia framework to global equity markets using price-based exposure networks.
    \item Integration of network topology and tail risk (heavy-tailed distributions) in systemic risk analysis.
    \item Quantification of cascade dynamics through deterministic and stochastic simulations.
    \item Identification of structural asymmetries between emerging and developed markets.
    \item Empirical evidence of localized contagion driven by clustered connectivity.
\end{itemize}

This study builds upon our previous work \cite{CastilloPereda2025}, where systemic risk and default cascades were initially analyzed using network-based approaches. The present paper extends that framework by incorporating stochastic simulations, threshold sensitivity analysis, and tail risk estimation, providing a more comprehensive and scalable methodology.

By integrating network science, stochastic modeling, and tail risk analysis, this framework advances systemic risk assessment in equity markets. The results offer practical insights for regulators and portfolio managers, supporting stress testing, risk monitoring, and diversification strategies in increasingly interconnected financial systems.
\section{Materials and Methods}

\subsection{Data and Risk Measures}

The dataset consists of daily low prices for 30 equity assets spanning the period from January 1, 2015, to April 2026, sourced from Yahoo Finance, a widely used and reliable public database for financial time series \cite{YahooFinance2023}. The sample includes a balanced selection of Brazilian and international assets from developed markets, chosen based on data availability, market capitalization, and sectoral diversity.

This configuration ensures analytical tractability while preserving sufficient structural complexity for network analysis, as larger networks often result in excessively dense graphs that obscure topological insights \cite{Newman2010}. The 30-asset structure enables clear identification of nodes and edges (e.g., Figures~\ref{fig:network_clustering_theta03} and \ref{fig:exposure_network_theta03}), allowing for a meaningful comparison between emerging and developed market dynamics.

Data preprocessing was conducted using Python, with \texttt{yfinance} for data acquisition, \texttt{pandas} and \texttt{numpy} for data manipulation, and \texttt{matplotlib} and \texttt{seaborn} for visualization. Missing observations were handled by excluding days with incomplete records and applying linear interpolation to address minor gaps, followed by forward and backward filling to ensure continuity of the time series.

To mitigate the impact of extreme outliers while preserving tail behavior, observations were filtered using the interquartile range (IQR) method, with bounds defined as $Q_1 - 1.5 \times \text{IQR}$ and $Q_3 + 1.5 \times \text{IQR}$, where $Q_1$ and $Q_3$ denote the first and third quartiles, respectively \cite{Tukey1977}.

The focus on daily low prices is intended to emphasize downside risk, aligning with the objective of capturing systemic vulnerability under adverse market conditions \cite{Cont2001, Das2023a, Das2023b}. Logarithmic returns were computed to standardize relative price variations across assets, as given by:
\begin{equation}
r_{i,t} = \ln\left(\frac{P_{i,t}}{P_{i,t-1}}\right),
\label{eq:log_returns}
\end{equation}
where $P_{i,t}$ denotes the low price of assets $i$ at time $t$.

Individual risk exposure was quantified using Value-at-Risk (VaR) and Conditional Value-at-Risk (CVaR) at the 95\% confidence level. VaR, representing the lower quantile of the return distribution, is defined as:
\begin{equation}
\text{VaR}_{i,\alpha} = F_i^{-1}(1 - \alpha),
\label{eq:var}
\end{equation}
where $F_i$ denotes the empirical cumulative distribution function of returns $r_{i,t}$, with $\alpha = 0.95$.

CVaR, capturing the expected loss conditional on exceeding the VaR threshold, is computed as:
\begin{equation}
\text{CVaR}_{i,\alpha} = \mathbb{E}[r_{i,t} \mid r_{i,t} \leq \text{VaR}_{i,\alpha}],
\label{eq:cvar}
\end{equation}
consistent with coherent risk measure frameworks \cite{Artzner1999}.

These tail risk measures play a central role in systemic risk analysis, as extreme losses in the lower tail of return distributions can act as triggers for cascading failures. By integrating VaR and CVaR with network-based contagion modeling, this study captures both individual asset vulnerability and the propagation of extreme events through interconnected financial structures.

\subsection{Network Construction}

An exposure-based financial network is constructed using an adapted Gai Kapadia framework. The correlation matrix $\rho$ of log returns is first computed, where asset volatility $\sigma_i$ is defined as the standard deviation of $r_{i,t}$.

Pairwise exposures between assets $i$ and $j$ are defined as:
\begin{equation}
E_{ij} = \rho_{ij} \cdot \sigma_i \cdot P_i,
\end{equation}
where $P_i$ represents a reference price level of asset $i$, used to scale exposures.
This formulation induces a directed weighted network, where exposures are asymmetric and depend on the originating asset's volatility and scale.

To extract the most relevant connections, exposures are filtered using thresholds $\theta \in \{0.3, 0.5\}$:
\begin{equation}
\tilde{E}_{ij} =
\begin{cases} 
E_{ij} & \text{if } E_{ij} \geq \theta, \\
0 & \text{otherwise}.
\end{cases}
\end{equation}

The filtered matrix $\tilde{E}_{ij}$ defines the adjacency structure of the network $G$, where nodes correspond to assets and edges represent significant exposure relationships.
This proxy captures relative exposure intensity based on co-movement and volatility, rather than balance-sheet linkages.

Local clustering coefficients are computed to quantify network connectivity:
\begin{equation}
C_i = \frac{2T_i}{k_i (k_i - 1)},
\end{equation}
where $T_i$ denotes the number of triangles involving node $i$, and $k_i$ is its degree.

The network is visualized using a force-directed (spring) layout, with nodes colored according to their clustering coefficients $C_i$, enabling structural comparison across different threshold levels and shock scenarios.

The thresholds $\theta \in \{0.3, 0.5\}$ are selected to balance network density and sparsity, ensuring meaningful connectivity while avoiding overly dense graphs that obscure cascade dynamics, a standard approach in financial network analysis \cite{Glasserman2016}.
These statistical properties not only characterize individual asset risk but also directly shape the structure of the financial network, as volatility and correlations determine the intensity of inter-asset exposures in the subsequent network construction.

\subsection{Default Cascade Model}

Default cascades are modeled using an adapted Gai Kapadia framework, implemented through both stochastic and deterministic approaches.

\textbf{Stochastic Simulations.} Each asset $i$ is assigned an initial capital $K_i = 0.2 \cdot P_i$ and a failure threshold $K_{\text{min},i} = 0.1 \cdot P_i$. A total of $n = 1000$ Monte Carlo simulations are performed, where each iteration applies a random shock $s \sim \text{Uniform}(0.1, 0.5)$, reducing the capital of affected assets.

Losses propagate through the network according to:
\begin{equation}
L_{ij} = \max\left(0, \tilde{E}_{ij} - (K_i - D_i)\right),
\end{equation}
where $D_i = \sum_j \tilde{E}_{ji}$ represents the total incoming exposure to asset $i$.

An asset $j$ defaults if its capital falls below the threshold, i.e., $K_j < K_{\text{min},j}$, triggering further contagion. The process iterates until no additional defaults occur. Systemic failure is defined as the collapse of more than five assets. The simulations produce measures including the probability of systemic failure, the average number of failed assets, and patterns of network fragility.

\textbf{Deterministic Propagation.} To analyze cascade dynamics, a deterministic version of the model is applied. An initial shock is introduced by setting a representative asset (e.g., $\texttt{VIVT3.SA}$) to default, and propagation is simulated iteratively.

For each asset $i$, the influence from defaulted neighbors is given by:
\begin{equation}
I_i = \sum_j \rho_{ij} \cdot S_j,
\end{equation}
where $S_j = 1$ if asset $j$ is in default and $0$ otherwise, and $\rho_{ij}$ is filtered by the threshold $\theta$.

Asset $i$ enters default if:
\begin{equation}
I_i > T_i,
\end{equation}
where $T_i$ is a fixed threshold (set to $0.5$ in this study).
The threshold $T_i = 0.5$ is chosen as a benchmark to represent a critical level of cumulative influence required to trigger default.

Default states are tracked across iterations to characterize cascade dynamics. Complementary evidence on the distribution of losses, including Pareto tail index estimates, is provided in the Appendix (Table~\ref{tab:pareto_tail_results}), supporting the role of extreme events in shaping systemic vulnerability.

\section{Results and Discussion}
\subsection{Evolution of Asset Prices and Descriptive Statistics}
\begin{figure}[H]
    \centering
    \includegraphics[width=0.8\textwidth]{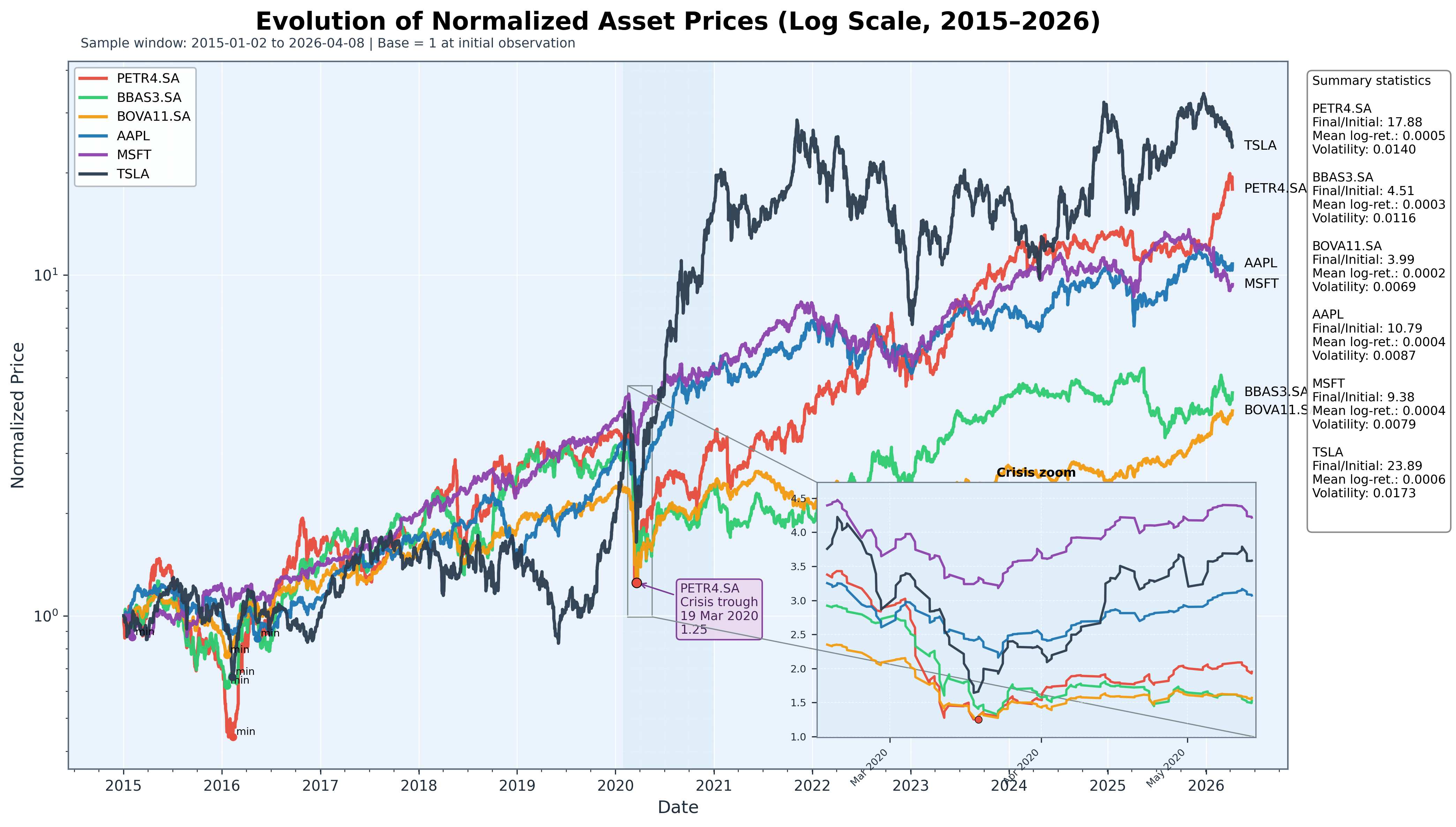}
    \caption{Evolution of normalized asset prices for selected Brazilian and developed-market assets over the period 2015–2026, displayed on a logarithmic scale. The shaded region highlights the 2020 market stress episode, while the inset zoom provides a detailed view of the crisis dynamics. Brazilian assets exhibit stronger drawdowns and higher dispersion during the shock, whereas developed-market assets show comparatively smoother recovery trajectories.}
    \label{fig:Assetnormalog}
\end{figure}

\textbf{Evolution of Normalized Asset Prices.} Figure~\ref{fig:Assetnormalog} depicts the evolution of normalized asset prices for selected Brazilian and international assets from 2015 to 2026. Prices are normalized as:
\begin{equation}
P_{\text{norm},i,t} = \frac{P_{i,t}}{P_{i,0}},
\end{equation}
where \(P_{i,t}\) is the price of asset \(i\) at time \(t\), and \(P_{i,0}\) is its initial price. Plotted on a logarithmic scale, the figure highlights stark contrasts in volatility and trends:
Brazilian assets such as \texttt{PETR4.SA} and \texttt{BBAS3.SA} exhibit pronounced drawdowns during the 2020 market stress episode, reflecting their higher sensitivity to external shocks such as COVID-19. This behavior is consistent with their strong intra-market correlations and elevated clustering coefficients (\(C_i \approx 0.7\text{--}1.0\)), as discussed in Section 3.2. 

In contrast, developed-market assets such as \texttt{AAPL} and \texttt{AMZN} display smoother trajectories and more stable recovery patterns, with lower volatility and reduced sensitivity to systemic disturbances (see Table~\ref{tab:log_returns_stats}). This difference highlights the greater resilience of developed markets, which is associated with lower effective connectivity (\(C_i \approx 0.2\text{--}0.4\)).

Overall, these dynamics illustrate the heterogeneous impact of systemic shocks across markets, providing visual support for the contagion mechanisms analyzed in Sections 3.3 and 3.4, where emerging-market assets exhibit higher susceptibility to cascade effects.
These findings reinforce the interpretation of financial markets as clustered networks, where densely connected regions amplify local shocks, leading to asymmetric propagation patterns. 

\textbf{Descriptive Statistics.} Table~\ref{tab:log_returns_stats} reports the descriptive statistics of log returns for the full set of assets. A clear distinction emerges between emerging and developed markets in terms of volatility and extreme returns. Brazilian assets such as \texttt{PETR4.SA} (Std. Dev. = 0.0140) and \texttt{MGLU3.SA} (Std. Dev. = 0.0190) exhibit higher volatility compared to developed-market assets such as \texttt{AAPL} (Std. Dev. = 0.0087) and \texttt{MSFT} (Std. Dev. = 0.0079), indicating greater sensitivity to market shocks.

Moreover, certain emerging-market assets display pronounced downside risk. For instance, \texttt{AMER3.SA} exhibits extreme negative returns (Min = -1.0553), suggesting exposure to idiosyncratic shocks and structural fragility. In contrast, developed-market assets tend to exhibit more concentrated return distributions and less severe extremes, reflecting higher market stability and liquidity~\cite{MantegnaStanley1999, Lux2016}.

These differences in volatility and extreme behavior are consistent with well-documented stylized facts in financial markets, where emerging economies typically exhibit heavier tails and higher kurtosis~\cite{Cont2001, Mandelbrot1963}. This asymmetry is particularly relevant for systemic risk analysis, as assets with higher volatility and more extreme losses contribute disproportionately to tail risk, as further evidenced by the VaR and CVaR measures reported in Table~\ref{tab:clustering_risk_measures}.

Importantly, these statistical properties directly inform the construction of the exposure-based network in Section 3.2. In particular, asset volatility (\(\sigma_i\)) enters the exposure metric \(E_{ij} = \rho_{ij} \cdot \sigma_i \cdot P_i\), amplifying connections associated with more volatile assets and increasing their potential to propagate shocks across the network.

Overall, the descriptive statistics highlight the heterogeneous risk structure across assets and markets, providing a quantitative foundation for the network-based systemic analysis developed in the subsequent sections. These patterns also support the interpretation that emerging-market assets play a central role in amplifying contagion dynamics under stress conditions.

This evidence reinforces the interpretation of financial markets as clustered systems, where volatility and extreme losses jointly determine the pathways of systemic risk transmission.

\begin{table}[H]
    \begin{center}
    \caption{Descriptive Statistics of Log Returns (2015 - 2026).}
    \label{tab:log_returns_stats}
    \begin{tabular}{l c c c c}\hline
    \textbf{Asset} & \textbf{Mean} & \textbf{Std. Dev.} & \textbf{Min} & \textbf{Max} \\
\hline
\hline
VIVT3.SA & 0.0003 & 0.0072 & -0.0569 & 0.0361 \\
PETR4.SA & 0.0005 & 0.0140 & -0.2084 & 0.1292 \\
ABEV3.SA & 0.0001 & 0.0075 & -0.0642 & 0.0654 \\
AMER3.SA & -0.0010 & 0.0308 & -1.0553 & 0.4155 \\
BBAS3.SA & 0.0003 & 0.0116 & -0.1507 & 0.0925 \\
BBDC4.SA & 0.0002 & 0.0099 & -0.1021 & 0.0615 \\
BOVA11.SA & 0.0002 & 0.0069 & -0.0804 & 0.0492 \\
RAIL3.SA & 0.0000 & 0.0149 & -0.2371 & 0.2202 \\
CSNA3.SA & 0.0001 & 0.0165 & -0.1710 & 0.0980 \\
ITUB4.SA & 0.0003 & 0.0086 & -0.1110 & 0.0528 \\
MGLU3.SA & 0.0003 & 0.0190 & -0.1420 & 0.1438 \\
VALE3.SA & 0.0004 & 0.0119 & -0.0984 & 0.1071 \\
WEGE3.SA & 0.0004 & 0.0093 & -0.0787 & 0.0887 \\
SUZB3.SA & 0.0002 & 0.0092 & -0.0864 & 0.1168 \\
LREN3.SA & 0.0001 & 0.0121 & -0.1666 & 0.0667 \\
AAPL & 0.0004 & 0.0087 & -0.0716 & 0.0606 \\
JPM & 0.0003 & 0.0085 & -0.1265 & 0.0921 \\
AMZN & 0.0005 & 0.0101 & -0.0740 & 0.0669 \\
MSFT & 0.0004 & 0.0079 & -0.0824 & 0.1001 \\
GOOGL & 0.0004 & 0.0085 & -0.0497 & 0.0738 \\
TSLA & 0.0006 & 0.0173 & -0.1114 & 0.1125 \\
V & 0.0003 & 0.0075 & -0.0874 & 0.0535 \\
SAP & 0.0002 & 0.0084 & -0.1396 & 0.0519 \\
NSRGY & 0.0001 & 0.0058 & -0.0434 & 0.0470 \\
SAN & 0.0001 & 0.0108 & -0.1220 & 0.0706 \\
HSBC & 0.0002 & 0.0078 & -0.0733 & 0.0571 \\
BABA & 0.0000 & 0.0129 & -0.1786 & 0.0902 \\
TM & 0.0001 & 0.0072 & -0.0519 & 0.0612 \\
SONY & 0.0003 & 0.0093 & -0.0672 & 0.0649 \\
HMC & 0.0000 & 0.0080 & -0.0753 & 0.0603 \\
\hline
    \end{tabular}
    \parbox{0.9\textwidth}{\vspace{0.3em} \small \textit{Note:} This table presents the mean, standard deviation, minimum, and maximum of daily log returns for 30 assets across Brazil and developed markets between 2015 and 2026.}
    \end{center}
\end{table}

\subsection{Network Structure and Clustering Analysis}

\begin{figure}[H]
    \centering
    \includegraphics[width=0.8\textwidth]{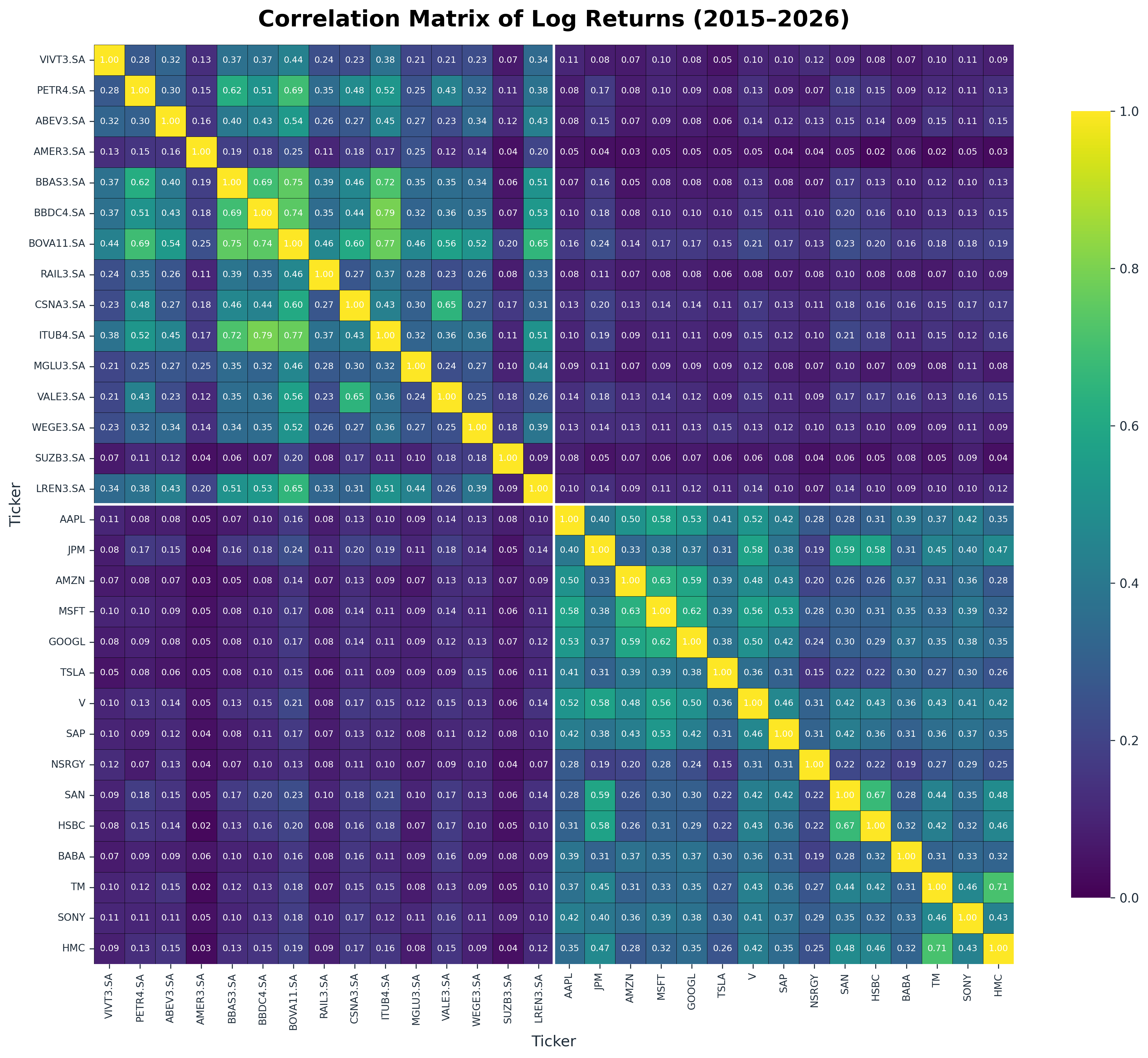}
   \caption{Correlation matrix of log returns for 30 assets (15 Brazilian and 15 developed-market assets) over the period 2015--2026. The heatmap reveals strong intra-market correlations among Brazilian equities and more moderate correlations among developed-market assets, while cross-market dependencies remain comparatively weaker. This clustered structure highlights the presence of localized connectivity patterns, with implications for diversification and the propagation of systemic risk.}
    \label{fig:corr_matrix}
\end{figure}

\textbf{Correlation Matrix.} Figure~\ref{fig:corr_matrix} presents the Pearson correlation matrix of log returns, defined as:
\begin{equation}
\rho_{ij} = \frac{\text{Cov}(R_i, R_j)}{\sigma_{R_i} \sigma_{R_j}},
\end{equation}
where \(\text{Cov}(R_i, R_j)\) denotes the covariance between asset returns, and \(\sigma_{R_i}\), \(\sigma_{R_j}\) are their respective standard deviations. 

The matrix reveals a clear block structure, with strong intra-market correlations among Brazilian assets often exceeding 0.6  and more moderate correlations among developed-market assets. For instance, financial institutions such as \texttt{BBAS3.SA}, \texttt{ITUB4.SA}, and \texttt{BBDC4.SA} exhibit consistently high pairwise correlations, indicating synchronized movements within the Brazilian market. In contrast, cross-market correlations between Brazilian and developed-market assets (e.g., \texttt{AAPL}, \texttt{MSFT}) remain relatively low, typically around 0.1–0.3, suggesting partial decoupling and potential diversification effects.

This correlation structure directly informs the construction of the exposure-based network in Section 3.2, where \(\rho_{ij}\) enters the exposure metric \(E_{ij} = \rho_{ij} \cdot \sigma_i \cdot P_i\). As a result, densely connected clusters—particularly within the Brazilian market—generate stronger effective links, increasing the likelihood of localized shock amplification.

These patterns are consistent with the cascade dynamics observed in Section 3.3, where shocks propagate more rapidly within highly correlated clusters, while transmission to less connected developed-market assets remains more limited. Consequently, the correlation matrix provides key evidence of a clustered network topology, in which tightly connected regions drive contagion, while weaker cross-market links contribute to partial systemic resilience.

This empirical structure aligns with network-based theories of systemic risk, where clustering and heterogeneity in connectivity play a central role in shaping contagion pathways.

\subsection{Deterministic Shock Analysis}

\begin{figure}[H]
    \centering
    \includegraphics[width=0.8\textwidth]{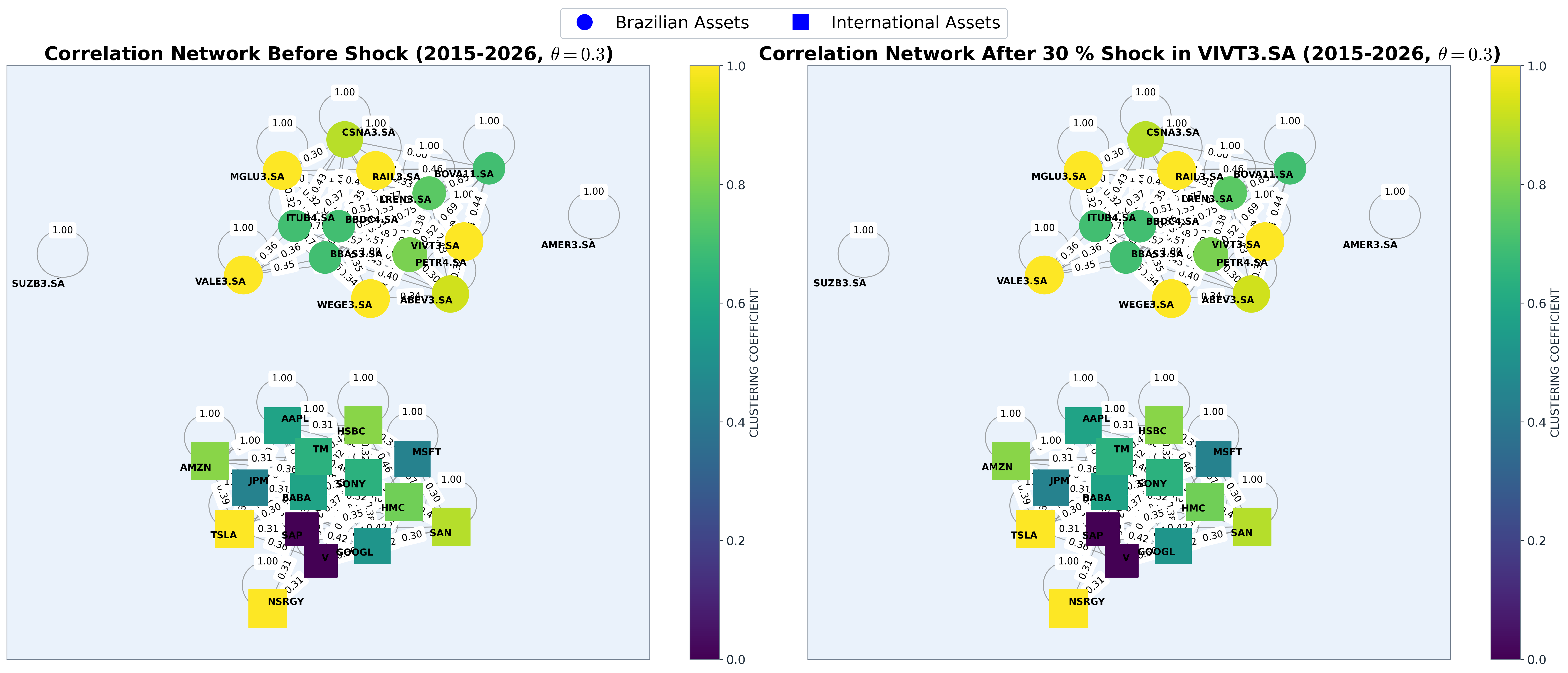}
   \caption{Correlation network before and after a 30\% negative shock applied to \texttt{VIVT3.SA} (\(2015 - 2026\), \(\theta = 0.3\)). The panels display correlation-based networks of Brazilian (circles) and developed-market assets (squares). Nodes are colored according to their clustering coefficients, with warmer colors indicating higher local connectivity. The shock induces noticeable adjustments in the topology of the Brazilian subnetwork, while the developed-market subnetwork remains comparatively stable, reflecting differences in connectivity and systemic resilience.}
    \label{fig:network_clustering_theta03}
\end{figure}

Figure~\ref{fig:network_clustering_theta03} compares the network before and after a deterministic 30\% negative shock applied to \texttt{VIVT3.SA}. The results show that the clustering coefficients of Brazilian assets exhibit noticeable adjustments, reflecting a reconfiguration of local connectivity within the cluster. 

Highly connected nodes such as \texttt{PETR4.SA}, \texttt{BBAS3.SA}, and \texttt{ITUB4.SA} - remain central to the network structure, but changes in edge intensities and clustering patterns indicate a redistribution of dependencies following the shock. In contrast, developed-market assets display minimal variation in their clustering coefficients, remaining largely unaffected by the localized disturbance.

These findings suggest that deterministic shocks primarily impact densely connected regions of the network, where higher clustering facilitates local propagation. Conversely, weaker connectivity in the developed-market subnetwork limits the transmission of shocks, contributing to greater structural stability.

This result reinforces the interpretation that clustering plays a central role in shaping shock propagation, acting as a local amplification mechanism within financial networks. 

\textbf{Threshold Sensitivity and Cascade Effects.} Figures~\ref{fig:corr_network_theta05} and \ref{fig:exposure_network_theta03}--\ref{fig:exposure_network_theta05} extend the network analysis by examining the impact of different correlation thresholds and exposure-based interactions on shock propagation.

At a higher threshold ($\theta = 0.5$), the correlation network becomes significantly sparser, retaining only the strongest connections. In this regime, the Brazilian subnetwork remains relatively cohesive, while the developed-market assets appear more fragmented. Following the shock to \texttt{VIVT3.SA}, only minor structural adjustments are observed, indicating that strong connections are more stable and less sensitive to localized perturbations.

In contrast, the exposure-based networks reveal a richer and more dynamic structure. Prior to the cascade, the network already exhibits heterogeneous connectivity, with certain assets acting as hubs due to their combined correlation and volatility profiles. After the cascade, the network undergoes a visible reconfiguration, with increased edge density and stronger connections emerging around key nodes.

Notably, the Brazilian assets continue to form a tightly connected core, while developed-market assets remain more peripherally connected. This pattern persists across both threshold levels, reinforcing the role of clustered structures in facilitating local contagion.

Overall, these results demonstrate that shock propagation is highly sensitive to both network construction and threshold selection. While higher thresholds isolate the most stable connections, exposure-based networks capture amplification effects more effectively, highlighting the importance of incorporating both correlation and volatility in systemic risk modeling.

\subsection{Default Cascades and Stochastic Simulations}
\begin{figure}[H]
    \centering
    \includegraphics[width=0.8\textwidth]{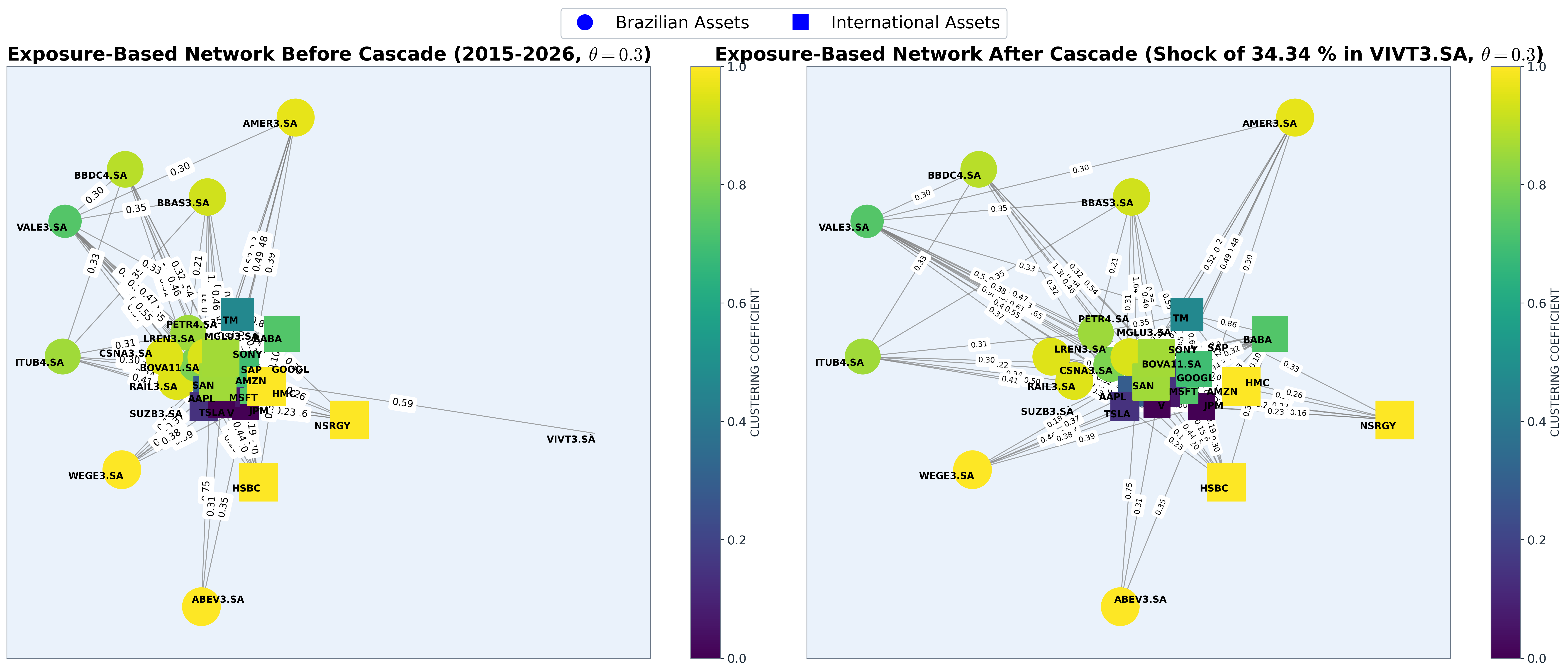}
    \caption{Exposure-based network before and after a default cascade (example simulation, \(\theta = 0.3\)).}
    \label{fig:exposure_network_theta03}
\end{figure}

Figure~\ref{fig:exposure_network_theta03} illustrates the network before and after a default cascade simulation, where a random shock to \texttt{VIVT3.SA} (10\% to 50\%) triggers loss propagation per the Gai-Kapadia mechanism. In this example, minimal node removal occurs, but clustering adjustments (\(\Delta C_{\text{VIVT3.SA}} \approx -0.05\)) indicate localized impacts.

\textbf{Network Structure Before and After Shock.} Figures~\ref{fig:network_clustering_theta03} and \ref{fig:exposure_network_theta03} depict the network structure with \(\theta = 0.5\) before and after a 30\% shock in \texttt{VIVT3.SA}. Brazilian assets (e.g., \texttt{BBAS3.SA}, \texttt{BOVA11.SA}) form a densely connected core (\(C_i \approx 0.8-1.0\)), while developed market assets (e.g., \texttt{AAPL}, \texttt{AMZN}) show lower connectivity (\(C_i \approx 0.0-0.4\)). Post-shock, the topology remains largely unchanged, reflecting structural resilience.

\begin{figure}[H]
    \centering
    \includegraphics[width=0.8\textwidth]{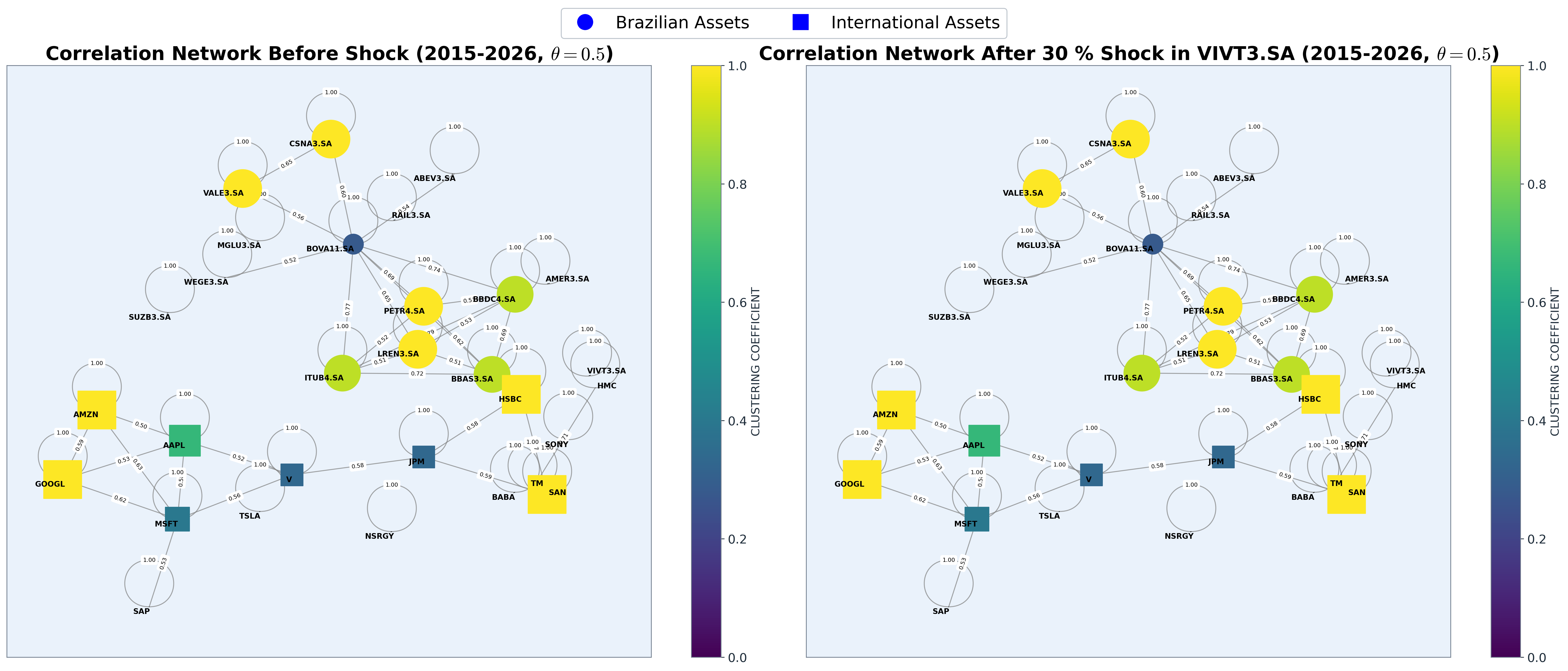}
    \caption{Network Structure Before Shock (2015-2026, \(\theta = 0.5\)), with Nodes Colored by Clustering Coefficient (\(C_i\)).}
    \label{fig:corr_network_theta05}
\end{figure}

\begin{figure}[H]
    \centering
    \includegraphics[width=0.8\textwidth]{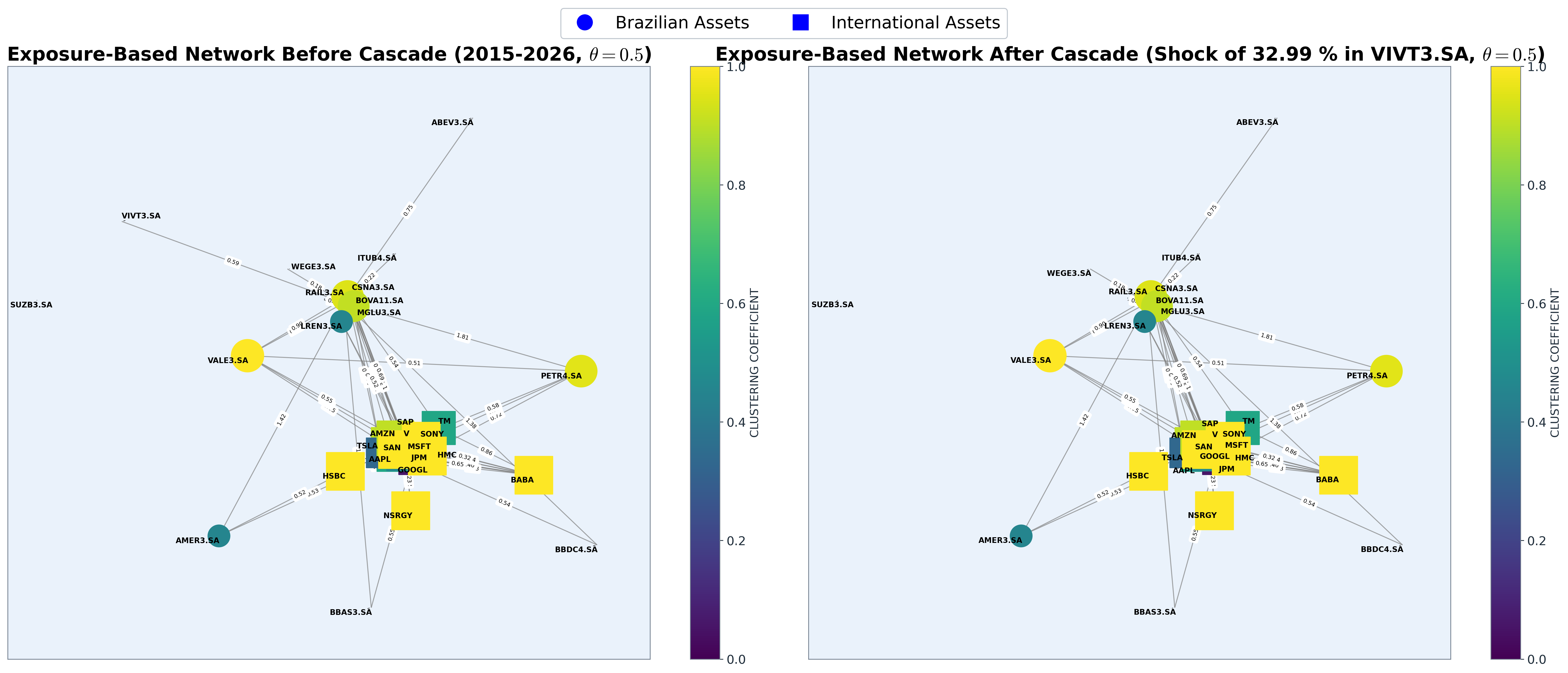}
    \caption{Network Structure After a 30\% Shock in \texttt{VIVT3.SA} (2015--2026, \(\theta = 0.5\)), with Nodes Colored by Clustering Coefficient (\(C_i\)).}
    \label{fig:exposure_network_theta05}
\end{figure}

\textbf{Relevance of Network Visualizations.} Figures~\ref{fig:network_clustering_theta03}, \ref{fig:corr_network_theta05}, and \ref{fig:exposure_network_theta03}-\ref{fig:exposure_network_theta05} collectively illustrate the structural dynamics of correlation and exposure-based networks under varying shock scenarios and thresholds (\(\theta = 0.3\) and \(\theta = 0.5\)).

These visualizations provide key insights into systemic risk propagation by revealing how network topology governs cascade dynamics. In particular, Brazilian assets consistently exhibit high clustering coefficients (\(C_i \approx 0.8\text{-}1.0\)) and significantly higher average degrees compared to developed-market assets, forming a dense and highly interconnected subnetwork. This structure facilitates rapid local shock transmission, as evidenced by the pronounced structural adjustments observed after shocks to \texttt{VIVT3.SA}, including reductions in clustering and reconfiguration of local connectivity.

In contrast, developed-market assets display lower clustering (\(C_i \approx 0.0\text{-}0.4\)) and sparser connectivity, resulting in more stable average degrees and limited structural changes following shocks. This relative sparsity acts as a buffer against contagion, restricting the propagation of disturbances across regions.

The exposure-based networks further emphasize these dynamics by incorporating both correlation and volatility into the link structure. Following cascade simulations, localized reductions in connectivity and edge intensities are observed around highly connected nodes, reflecting the sensitivity of the network to exposure thresholds in the Gai Kapadia framework.

Overall, the combined analysis of clustering coefficients, average degree, and edge dynamics provides a robust quantitative framework for assessing systemic vulnerabilities. These results align with established findings in financial network literature~\cite{Gai2010, Glasserman2016}, highlighting the critical role of clustered connectivity in amplifying contagion and shaping the resilience of financial systems. Such insights are particularly relevant for identifying systemically important nodes and informing risk management and diversification strategies in interconnected markets.

\textbf{Clustering Coefficients.} Table~\ref{tab:clustering_risk_measures} reports clustering coefficients for $\theta = 0.3$ and $\theta = 0.5$ alongside tail risk measures. The results confirm that Brazilian assets exhibit consistently higher clustering (e.g., \texttt{PETR4.SA}, $C_i=0.855$ at $\theta=0.3$ and $0.700$ at $\theta=0.5$; \texttt{BBAS3.SA}, $C_i=0.929$ at $\theta=0.3$), whereas developed-market assets display more moderate values (e.g., \texttt{JPM}, $C_i=0.473$; \texttt{MSFT}, $C_i=0.505$ at $\theta=0.3$), despite some highly connected nodes (e.g., \texttt{NSRGY}, $C_i=1.000$).

This disparity highlights a more densely interconnected Brazilian subnetwork, in which high $C_i$ values reflect tightly clustered local structures that facilitate rapid intra-cluster transmission of shocks. Consistently, under deterministic shocks (Section~3.3), perturbations to \texttt{VIVT3.SA} lead to observable reconfigurations within this cluster, with changes in clustering and edge intensities indicating localized amplification.

Moreover, when combined with tail risk measures, highly clustered assets tend to exhibit larger downside risk (e.g., \texttt{AMER3.SA}, $C_i=0.964$ at $\theta=0.3$, CVaR $=-0.0633$), reinforcing the link between network connectivity and vulnerability to extreme events. Overall, these findings support the interpretation that clustering plays a central role in shaping contagion dynamics, acting as a local amplification mechanism within financial networks.

\begin{table}[H]
\caption{Clustering Coefficients and Risk Measures of Assets (2015-2026).}
\label{tab:clustering_risk_measures}
\begin{center}
\begin{tabular}{lcccc}\hline
\textbf{Asset} & \textbf{\(\theta = 0.3\)} & \textbf{\(\theta = 0.5\)} & \textbf{VaR (95\%)} & \textbf{CVaR (95\%)} \\\hline
VIVT3.SA & 0.000 & 0.000 & -0.0110 & -0.0169 \\
PETR4.SA & 0.855 & 0.700 & -0.0191 & -0.0336 \\
ABEV3.SA & 1.000 & 0.000 & -0.0108 & -0.0179 \\
AMER3.SA & 0.964 & 0.333 & -0.0302 & -0.0633 \\
BBAS3.SA & 0.929 & 0.000 & -0.0164 & -0.0267 \\
BBDC4.SA & 0.893 & 0.000 & -0.0143 & -0.0232 \\
BOVA11.SA & 0.407 & 0.026 & -0.0098 & -0.0156 \\
RAIL3.SA & 0.952 & 0.000 & -0.0189 & -0.0361 \\
CSNA3.SA & 0.802 & 0.694 & -0.0249 & -0.0367 \\
ITUB4.SA & 0.857 & 0.000 & -0.0120 & -0.0193 \\
MGLU3.SA & 0.944 & 0.667 & -0.0282 & -0.0446 \\
VALE3.SA & 0.733 & 0.733 & -0.0178 & -0.0269 \\
WEGE3.SA & 1.000 & 0.000 & -0.0132 & -0.0213 \\
SUZB3.SA & 0.000 & 0.000 & -0.0133 & -0.0211 \\
LREN3.SA & 0.952 & 0.333 & -0.0173 & -0.0276 \\
AAPL & 0.626 & 0.769 & -0.0135 & -0.0210 \\
JPM & 0.473 & 0.468 & -0.0124 & -0.0204 \\
AMZN & 0.857 & 0.945 & -0.0150 & -0.0243 \\
MSFT & 0.505 & 0.437 & -0.0121 & -0.0191 \\
GOOGL & 0.547 & 0.705 & -0.0132 & -0.0205 \\
TSLA & 0.547 & 0.625 & -0.0263 & -0.0422 \\
V & 0.473 & 0.480 & -0.0115 & -0.0193 \\
SAP & 0.838 & 0.945 & -0.0122 & -0.0202 \\
NSRGY & 1.000 & 1.000 & -0.0084 & -0.0135 \\
SAN & 0.924 & 1.000 & -0.0162 & -0.0258 \\
HSBC & 1.000 & 1.000 & -0.0121 & -0.0195 \\
BABA & 0.857 & 1.000 & -0.0193 & -0.0301 \\
TM & 0.719 & 0.769 & -0.0112 & -0.0168 \\
SONY & 0.927 & 1.000 & -0.0136 & -0.0218 \\
HMC & 1.000 & 1.000 & -0.0127 & -0.0185 \\
\end{tabular}
\parbox{0.9\textwidth}{\small\textit{Note:} Clustering coefficients for \(\theta = 0.3\) reflect the denser network structure at this threshold, leading to more uniform values for developed market assets due to lower connectivity. VaR and CVaR (95\%) are calculated based on daily returns from 2015 to 2026.}
\end{center}
\end{table}
 In contrast, the lower clustering of developed market assets reflects a more fragmented network structure, contributing to their resilience against contagion, as evidenced by the absence of failures beyond Brazilian assets in stochastic simulations (Section 3.4).
 
\subsection{Risk Measures}

Table~\ref{tab:clustering_risk_measures} reports VaR and CVaR at the 95\% confidence level alongside clustering coefficients. The results highlight higher tail risks among several emerging-market assets (e.g., \texttt{AMER3.SA}, CVaR $=-0.0633$; \texttt{MGLU3.SA}, CVaR $=-0.0446$) compared to developed-market assets (e.g., \texttt{AAPL}, CVaR $=-0.0210$; \texttt{MSFT}, CVaR $=-0.0191$), underscoring heterogeneous exposure to extreme losses.

These risk measures provide critical insights into downside vulnerability and, when combined with the elevated clustering observed in Brazilian assets ($C_i \approx 0.8\text{--}1.0$), point to an amplification mechanism within densely connected subnetworks. In such structures, extreme losses are more likely to propagate locally, increasing systemic impact.

Consistent with the deterministic cascade analysis (Section~3.3), shocks to \texttt{VIVT3.SA} lead to localized reconfigurations and propagation within the Brazilian cluster, while the comparatively lower tail risks and weaker connectivity of developed-market assets limit contagion effects in both deterministic and stochastic settings (Sections~3.3 and 3.4).

Overall, the joint evidence from tail risk measures and network topology supports the view that systemic risk arises from the interaction between extreme loss potential and clustered connectivity, rather than from isolated asset characteristics alone.

\subsection{Stochastic and Deterministic Cascade Analysis}

\textbf{Stochastic Simulations.} Table~\ref{tab:monte_carlo_results} summarizes the results of the Monte Carlo simulations ($n = 1000$). Across all scenarios and thresholds ($\theta = 0.3$ and $\theta = 0.5$), the probability of systemic failure (defined as more than 5 assets failing) remains equal to zero, indicating the absence of large-scale collapse under the simulated conditions.

Regardless of the shock configuration—whether general, single-asset (e.g., \texttt{ITUB4.SA}), or simultaneous shocks (e.g., \texttt{ITUB4.SA} + \texttt{VALE3.SA})—the average number of failed assets remains stable, equal to 1.000 for isolated shocks and 2.000 for simultaneous shocks. This consistency suggests that the system exhibits localized vulnerability, where shocks produce limited propagation confined to a small subset of nodes.

These findings indicate that, while the network is susceptible to localized disturbances, it operates below a critical threshold required for systemic cascades. This behavior is consistent across both threshold levels, highlighting the robustness of the system under moderate shock intensities.

Overall, the results support the interpretation that systemic risk in this setting manifests through contained, localized failures rather than widespread collapse, in line with the predictions of the Gai--Kapadia framework.

\begin{table}[H]
    \begin{center}
    \caption{Monte Carlo Simulation Results ($n = 1000$) under different shock scenarios and thresholds.}
    \label{tab:monte_carlo_results}
    \begin{tabular}{l c c c}\hline
        \textbf{Scenario/Metric} & \textbf{$\theta$} & \textbf{Failure Probability ($>5$ assets)} & \textbf{Avg. Failed Assets} \\\hline
        General Simulation & 0.3 & 0.000 & 1.000 \\
        General Simulation & 0.5 & 0.000 & 1.000 \\
        Single Shock (VIVT3.SA) & 0.3 & 0.000 & 1.000 \\
        Single Shock (VIVT3.SA) & 0.5 & 0.000 & 1.000 \\
        Simultaneous Shock (VIVT3.SA + AAPL) & 0.3 & 0.000 & 2.000 \\
        Simultaneous Shock (VIVT3.SA + AAPL) & 0.5 & 0.000 & 2.000 \\
    \hline
    \end{tabular}
    \end{center}
\end{table}

\begin{figure}[H]
    \centering
    \includegraphics[width=0.8\textwidth]{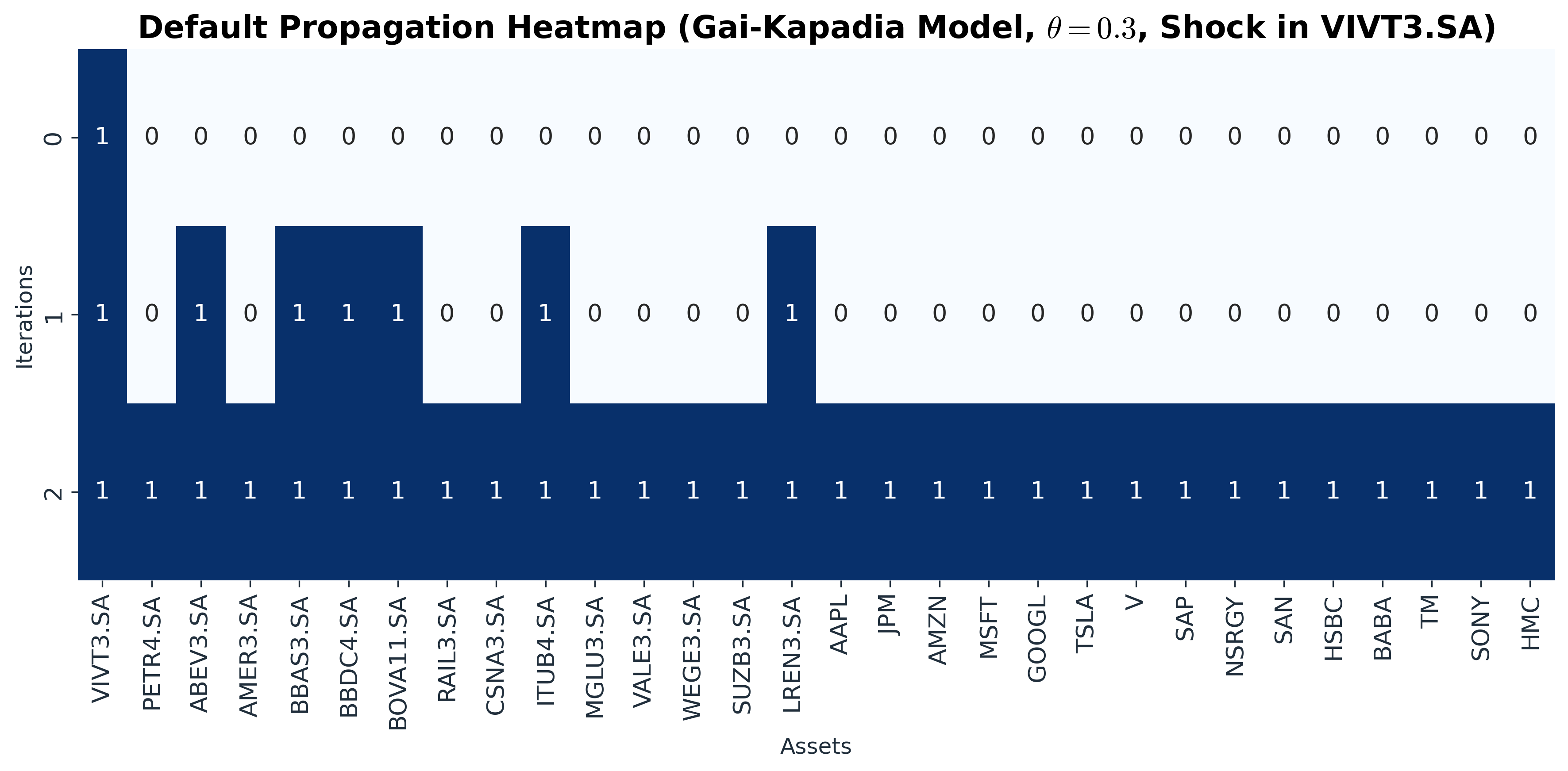}
 \caption{Default propagation heatmap under the Gai-Kapadia model ($\theta = 0.3$), following a shock to \texttt{VIVT3.SA}. The figure illustrates the evolution of default states across iterations, highlighting the progressive spread of failures within the Brazilian subnetwork and the absence of contagion in developed-market assets.}
    \label{fig:default_propagation_theta03}
\end{figure}

\textbf{Default Propagation Dynamics.} To further investigate the temporal evolution of contagion, we analyze the propagation of defaults across iterations within the Gai Kapadia framework. Figure~\ref{fig:default_propagation_theta03} presents a heatmap representation of default states, where each row corresponds to an iteration step and each column represents an asset.

The heatmap reveals a rapid transition from an initial localized shock to a stable configuration in which failures remain confined to a subset of assets. Notably, the propagation stabilizes after a small number of iterations, indicating that the system reaches a steady state without triggering a full cascade.

Formally, the default dynamics follow the recursive condition:
\begin{equation}
D_i^{(t+1)} = \mathbb{I} \left( \sum_{j} E_{ij} D_j^{(t)} > \tau_i \right),
\end{equation}
where $D_i^{(t)}$ denotes the default state of asset $i$ at iteration $t$, $E_{ij}$ represents the exposure matrix, and $\tau_i$ is the default threshold.

This result confirms that contagion is driven by local network structure and exposure intensity, rather than global interdependence.

\textbf{Tail Risk Analysis.} To quantify extreme loss behavior, we examine the empirical complementary cumulative distribution function (CCDF) of losses for selected assets.

\begin{figure}[H]
    \centering
    \includegraphics[width=0.8\textwidth]{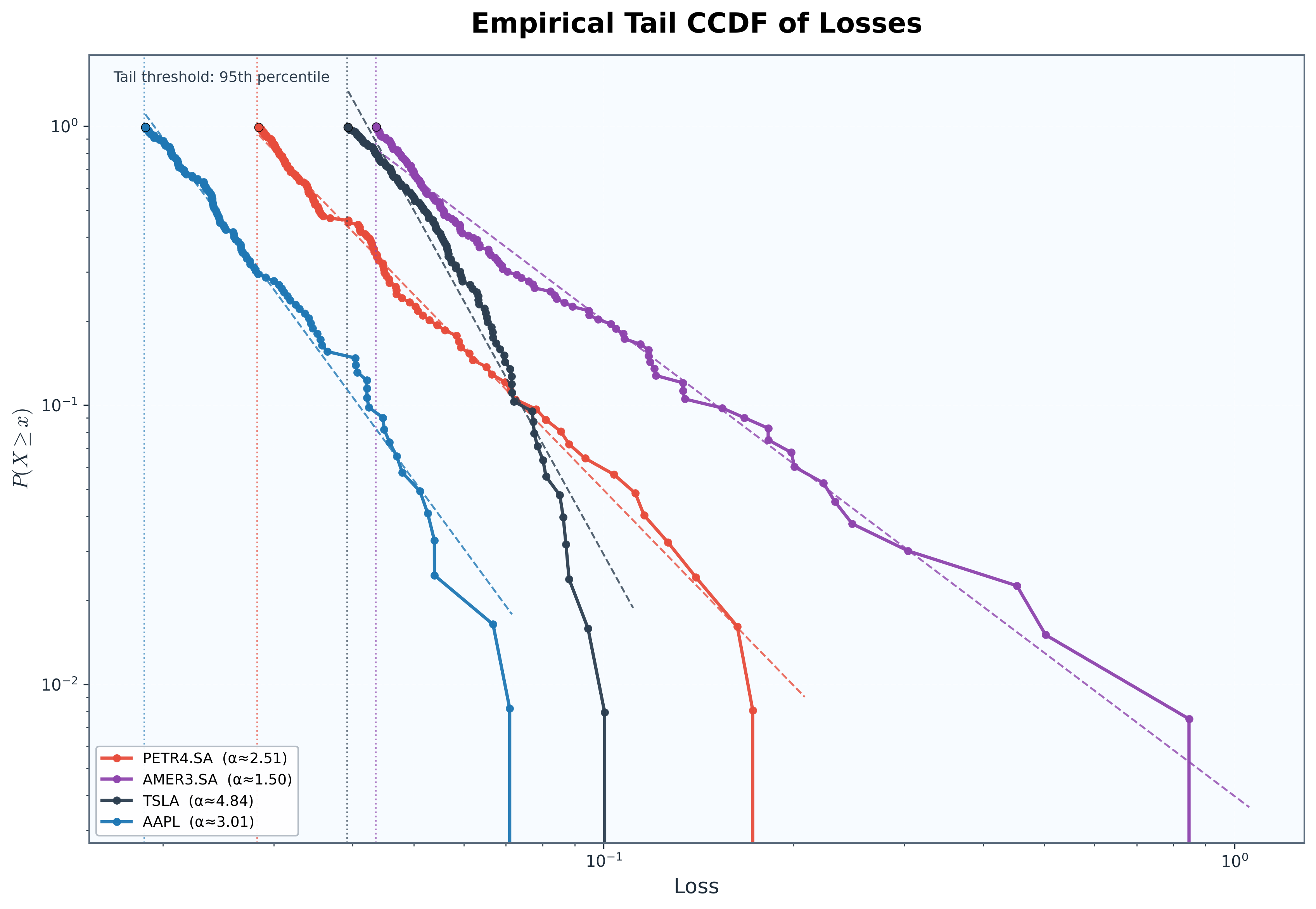}
  \caption{Empirical tail CCDF of losses for selected assets. The dashed lines represent power-law fits above the 95th percentile threshold, highlighting differences in tail behavior across assets. Brazilian assets exhibit heavier tails compared to developed-market assets, indicating higher exposure to extreme losses.}
    \label{fig:tail}
\end{figure}

The CCDF exhibits approximately linear behavior in log-log scale, suggesting a power-law decay of the form:
\begin{equation}
P(X \geq x) \sim x^{-\alpha},
\end{equation}
where $\alpha$ is the Pareto tail index.

Assets such as \texttt{AMER3.SA} display significantly heavier tails ($\alpha \approx 1.5$), indicating a higher probability of extreme losses, while developed-market assets exhibit steeper decay and greater stability.

These findings reinforce the link between tail risk and systemic vulnerability, particularly when combined with high clustering in the network structure.

\textbf{Tail Index Estimation.} To assess the robustness of the tail behavior, we estimate the Pareto tail index using the Hill estimator.

\begin{figure}[H]
    \centering
    \includegraphics[width=0.8\textwidth]{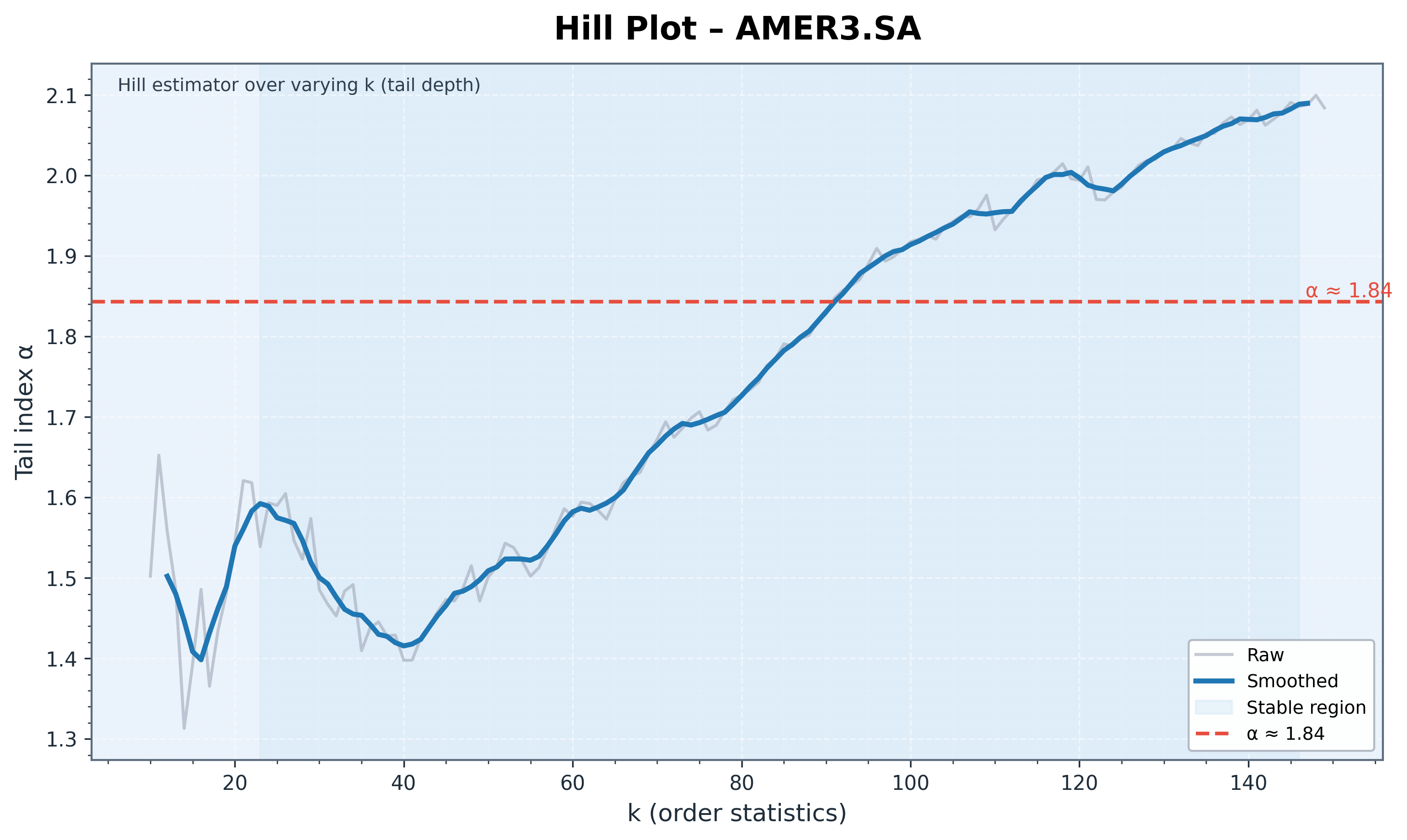}
  \caption{Hill plot for \texttt{AMER3.SA}, showing the estimated Pareto tail index $\alpha$ as a function of the number of upper order statistics $k$. The shaded region indicates the stable estimation interval, with $\alpha \approx 1.84$.}
    \label{fig:hill}
\end{figure}
The Hill estimator is defined as:
\begin{equation}
\hat{\alpha}^{-1}(k) = \frac{1}{k} \sum_{i=1}^{k} \left( \log X_{(i)} - \log X_{(k+1)} \right),
\end{equation}
where $X_{(i)}$ denotes the ordered sample.

The plot reveals a stable region for intermediate values of $k$, supporting the presence of a heavy-tailed distribution. The estimated tail index ($\alpha \approx 1.84$) confirms that extreme events are significantly more likely than under Gaussian assumptions.

This behavior is consistent with the CCDF results and provides further evidence that tail risk plays a central role in systemic vulnerability.

The evolution of defaults across iterations (Figure~\ref{fig:default_propagation_theta03}) provides further insight into the mechanisms of contagion within the network. Starting from an initial shock to \texttt{VIVT3.SA}, failures propagate rapidly to a subset of Brazilian assets within the first iteration, reflecting strong interdependence driven by high correlation and exposure levels.

By the second iteration, the system reaches a stable configuration in which defaults remain confined to a localized cluster, with no further propagation across the global network. This behavior indicates that, although contagion is present, it is limited by the network structure and does not escalate into a systemic cascade.

In contrast, developed-market assets (e.g., \texttt{AAPL}) remain unaffected throughout all iterations, highlighting their lower connectivity and reduced exposure to contagion channels. This asymmetry reinforces the role of network topology in shaping financial stability, where densely connected subnetworks act as amplifiers of local shocks, while sparse structures provide resilience.

\textbf{Key Findings:}
\begin{itemize}
    \item Brazilian assets exhibit higher clustering and stronger local contagion effects.
    \item Developed markets remain largely insulated due to weaker connectivity.
    \item Default propagation stabilizes after a few iterations, preventing systemic collapse.
    \item Tail risk and network structure jointly determine systemic vulnerability.
\end{itemize}

The empirical findings presented above provide a consistent picture of how network topology and tail behavior jointly shape systemic risk, which we now interpret in a broader financial and theoretical context.

\section{Discussion}

This study integrates deterministic cascade modeling and stochastic simulations within an extended Gai--Kapadia framework to analyze systemic risk in a 30-asset global equity network over the period 2015--2026. The results provide consistent evidence that systemic risk is primarily shaped by the interaction between network topology and tail risk characteristics.

The network analysis reveals a clear structural asymmetry between emerging and developed markets. Brazilian assets exhibit high clustering coefficients ($C_i \approx 0.8$--$1.0$) and dense connectivity, forming tightly interconnected subnetworks. In contrast, developed-market assets display lower clustering ($C_i \approx 0.2$--$0.5$) and sparser connections, acting as structural buffers against contagion.

Deterministic cascade results show that shocks propagate rapidly within the Brazilian subnetwork but stabilize after a few iterations, without triggering global systemic collapse. This behavior is further confirmed by stochastic simulations ($n = 1000$), which yield a zero probability of large-scale failure (defined as more than five assets defaulting), while maintaining a consistent average of failed assets (1.0 for single shocks and 2.0 for simultaneous shocks). These findings indicate a system operating below a critical percolation threshold.

Importantly, tail risk analysis provides an additional dimension to systemic vulnerability. The empirical CCDF and Hill estimator confirm the presence of heavy-tailed loss distributions, particularly in emerging-market assets (e.g., $\alpha \approx 1.5$--$2.0$), implying a higher likelihood of extreme events compared to developed markets. This heavy-tailed behavior, when combined with high clustering, creates a structural amplification mechanism for localized contagion.

The results highlight that systemic risk does not arise solely from extreme losses or network connectivity in isolation, but from their interaction. Highly connected assets with elevated tail risk emerge as critical nodes capable of amplifying shocks within the system.

Despite these insights, the analysis abstracts from liquidity effects, which are known to play a central role during financial crises. The absence of liquidity-driven feedback mechanisms (e.g., fire sales, funding constraints) may lead to an underestimation of systemic risk under extreme scenarios. Additionally, the use of static thresholds ($\theta = 0.3, 0.5$) simplifies the dynamic nature of financial correlations.

Future research could address these limitations by incorporating adaptive thresholds, liquidity constraints, and multi-shock scenarios, as well as extending the framework to larger and more heterogeneous networks. Nevertheless, the present framework provides a robust and scalable approach for understanding the structural drivers of systemic risk in equity markets.

\section{Conclusion}

This study develops a comprehensive framework for analyzing systemic risk and default cascades in global equity markets using an extended Gai Kapadia model. By combining network topology, stochastic simulations, and tail risk analysis, the results provide a unified view of how systemic risk emerges and propagates.

The findings demonstrate that the financial system exhibits strong resilience at the global level, with negligible probability of large-scale collapse. However, localized vulnerability persists within highly clustered subnetworks, particularly among emerging-market assets. This duality highlights that systemic risk is not uniformly distributed, but instead concentrated in structurally dense regions of the network.

Moreover, the integration of tail risk measures reveals that extreme events play a crucial role in shaping systemic outcomes. Assets with heavy-tailed loss distributions and high connectivity act as key drivers of contagion, reinforcing the importance of jointly considering statistical and topological properties.

These results extend the applicability of the Gai Kapadia framework beyond interbank networks, demonstrating its effectiveness in capturing price-driven contagion in equity markets. From a policy perspective, the identification of highly clustered and high-risk nodes provides a basis for targeted regulatory interventions. For portfolio management, the findings emphasize the importance of diversification across structurally distinct regions.

Future work may enhance this framework by incorporating dynamic network structures, liquidity effects, and more complex shock distributions, further bridging the gap between theoretical modeling and real-world financial systems.

Overall, this study contributes to the growing literature on financial networks by providing a scalable and empirically grounded approach to systemic risk analysis, offering insights relevant for both academic research and practical risk management.

\section*{Author Contributions}
Ana Isabel Castillo Pereda contributed to the conceptualization, methodology, data analysis, and writing of the manuscript, including both the original draft and subsequent revisions.

\appendix

\subsection{Tail Risk Estimates}

\begin{table}[H]
\caption{Pareto Tail Index Estimates and Loss Statistics for All Assets.}
\label{tab:pareto_tail_results}
\begin{center}
\begin{tabular}{lccc}
\toprule
\textbf{Asset} & \textbf{Pareto $\alpha$} & \textbf{$N_{\text{losses}}$} & \textbf{Tail Type} \\
\midrule
AMER3.SA & 1.501 & 2643 & Heavy tail \\
BBAS3.SA & 1.862 & 2555 & Heavy tail \\
BOVA11.SA & 2.067 & 2576 & Heavy tail \\
RAIL3.SA & 2.195 & 2547 & Heavy tail \\
BBDC4.SA & 2.230 & 2698 & Heavy tail \\
JPM & 2.246 & 2527 & Heavy tail \\
ITUB4.SA & 2.258 & 2601 & Heavy tail \\
SAP & 2.388 & 2569 & Heavy tail \\
WEGE3.SA & 2.479 & 2556 & Heavy tail \\
PETR4.SA & 2.508 & 2475 & Heavy tail \\
SAN & 2.605 & 2573 & Heavy tail \\
LREN3.SA & 2.629 & 2697 & Heavy tail \\
HMC & 2.687 & 2736 & Heavy tail \\
HSBC & 2.768 & 2458 & Heavy tail \\
CSNA3.SA & 2.875 & 2785 & Heavy tail \\
MSFT & 2.971 & 2480 & Heavy tail \\
\midrule
AAPL & 3.013 & 2438 & Moderate tail \\
VIVT3.SA & 3.044 & 2610 & Moderate tail \\
NSRGY & 3.096 & 2651 & Moderate tail \\
V & 3.096 & 2453 & Moderate tail \\
ABEV3.SA & 3.166 & 2632 & Moderate tail \\
TM & 3.195 & 2640 & Moderate tail \\
MGLU3.SA & 3.210 & 2876 & Moderate tail \\
SUZB3.SA & 3.278 & 2072 & Moderate tail \\
BABA & 3.326 & 2709 & Moderate tail \\
VALE3.SA & 3.358 & 2653 & Moderate tail \\
AMZN & 3.463 & 2512 & Moderate tail \\
SONY & 3.796 & 2652 & Moderate tail \\
GOOGL & 3.816 & 2490 & Moderate tail \\
TSLA & 4.844 & 2512 & Moderate tail \\
\bottomrule
\end{tabular}
\end{center}
\end{table}

Table~\ref{tab:pareto_tail_results} reports the estimated Pareto tail indices for all assets. 
Assets with $\alpha < 3$ exhibit heavy-tailed behavior, indicating higher exposure to extreme losses, 
particularly among Brazilian equities.
\end{document}